\documentclass[aps,prl,preprint,superscriptaddress]{revtex4}

\usepackage{graphicx}
\usepackage{color}

\begin{document}
\title{Coexistence of the topological state and a two-dimensional electron gas on the surface of Bi$_2$Se$_3$}
\author{Marco Bianchi}
\affiliation{Department of Physics and Astronomy, Interdisciplinary Nanoscience Center, Aarhus University,
8000 Aarhus C, Denmark}
\author{Dandan Guan}
\affiliation{Department of Physics and Astronomy, Interdisciplinary Nanoscience Center, Aarhus University,
8000 Aarhus C, Denmark}
\affiliation{Department of Physics, Zhejiang University, Hangzhou, 310027 China}
\author{Shining Bao}
\affiliation{Department of Physics, Zhejiang University, Hangzhou, 310027 China}
\author{Jianli Mi}
\affiliation{Center for Materials Crystallography, Department of Chemistry, Interdisciplinary Nanoscience Center, Aarhus University,
8000 Aarhus C, Denmark}
\author{Bo Brummerstedt Iversen}
\affiliation{Center for Materials Crystallography, Department of Chemistry, Interdisciplinary Nanoscience Center, Aarhus University,
8000 Aarhus C, Denmark}
\author{Philip~D.~C.~King}
\affiliation{School of Physics and Astronomy, University of St.\ Andrews, St.\ Andrews, KY16 9SS, United Kingdom}
\author{Philip Hofmann}
\affiliation{Department of Physics and Astronomy, Interdisciplinary Nanoscience Center, Aarhus University,
8000 Aarhus C, Denmark}
\email[]{philip@phys.au.dk}

\date{\today}
%
%

\maketitle

\textbf{Topological insulators are a recently discovered class of materials with fascinating properties: While the inside of the solid is insulating, fundamental symmetry considerations require the surfaces to be metallic \cite{Fu:2007b,Fu:2007c,Moore:2007,Hsieh:2008,Zhang:2008,Moore:2010}. The metallic surface states show an unconventional spin texture  \cite{Hsieh:2009,Hsieh:2009b}, electron dynamics  \cite{Pascual:2004,Konig:2007,Roushan:2009,Alpichshev:2010} and stability. Recently, surfaces with only a single Dirac cone dispersion have received particular attention \cite{Hsieh:2009b,Hsieh:2009c,Xia:2009,Zhang:2009,Chen:2009,Lin:2010a,Lin:2010b}. These are predicted to play host to a number of novel physical phenomena such as Majorana fermions  \cite{Fu:2008}, magnetic monopoles \cite{Qi:2009b} and unconventional superconductivity \cite{Lindner:2010}. Such effects will mostly occur when the topological surface state lies in close proximity to a magnetic or electric field, a (superconducting) metal, or if the material is in a confined geometry. Here we show that a band bending near to the surface of the topological insulator Bi$_2$Se$_3$ gives rise to the formation of a two-dimensional electron gas (2DEG). The 2DEG, renowned from semiconductor surfaces and interfaces where it forms the basis of the integer and fractional quantum Hall effects  \cite{Klitzing:1980,Tsui:1982}, two-dimensional superconductivity \cite{Reyren:2007}, and a plethora of practical applications \cite{Ando:1982}, coexists with the topological surface state in Bi$_2$Se$_3$. This leads to the unique situation where a topological and a non-topological, easily tunable and potentially superconducting, metallic state are confined to the same region of space. 
 }

\begin{figure}
\begin{center}
\includegraphics[width=\textwidth]{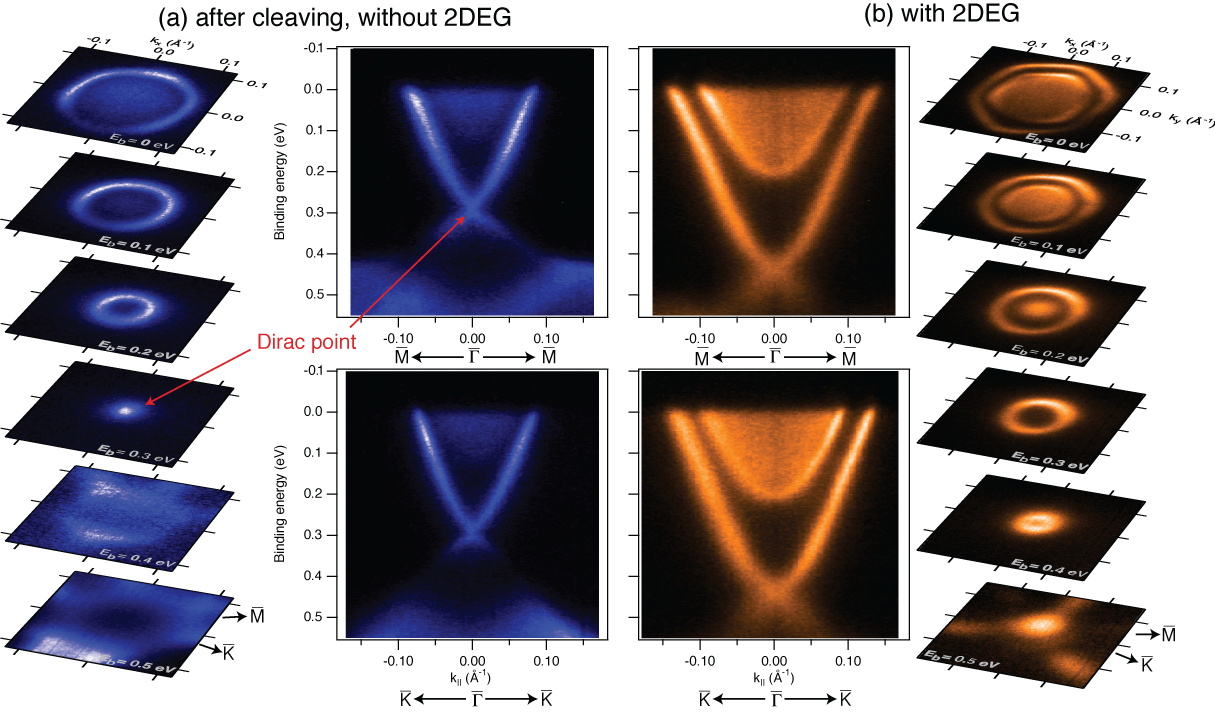}
\caption{{\bf Development of a surface 2DEG in Bi$_2$Se$_3$.} (a) Electronic structure of  Bi$_2$Se$_3$(0001), around 30 minutes after cleaving the sample, presented as cuts at different binding energies ($E_b$) and dispersions along high-symmetry directions. The sharp V-like feature is the Dirac-like dispersion of the topological surface state. The valence band states are visible below the Dirac point and the diffuse intensity inside the Dirac cone is due to the filled bottom of the conduction band. (b) After three hours in vacuum, the chemical potential of the material is shifted, presumably due to the formation of defects. This shift is accompanied by the emergence of a sharp rim around the bottom of the conduction band. This is the spectral signature of the 2DEG near the surface. The constant energy contours of the topological state, the 2DEG, and the conduction band show a hexagonal shape near the Fermi energy. \label{fig:1}}
\end{center}
\end{figure}

The emergence of a 2DEG on the (0001) surface of  Bi$_2$Se$_3$ can be followed by angle-resolved photoemission (ARPES), as shown in Fig. \ref{fig:1}. Immediately after cleaving the sample, a sharp V-shaped topological state is observed in the gap. Features stemming from the bulk valence and conduction bands are also seen. The latter should not be observable by photoemission for an intrinsic semiconductor, but most as-grown Bi$_2$Se$_3$ crystals turn out to be degenerately electron-doped  \cite{Urazhdin:2004,Hsieh:2009c}. Over a period of a few hours (depending on the sample temperature), spectral changes are observed: The chemical potential shifts gradually upwards, an effect which has been ascribed to a defect-induced downward band bending near the surface \cite{Hsieh:2009c}. Eventually, we find that an intense and narrow rim emerges around the bottom of the conduction band (Fig. \ref{fig:1}(b)). This state is assigned to a quantum-confined 2DEG, similar to what has been found for other narrow-gap semiconductors \cite{Piper:2008,King:2010}. After about three hours the situation becomes more stable, with the sample gradually deteriorating but no further pronounced shifts of the chemical potential. 

Fig. \ref{fig:1}(b) also shows the dispersion of the topological state and the two-dimensional electron gas as constant energy cuts through a three-dimensional data set. The topological state has a circular constant energy contour near the Dirac point, but all the states (topological state, 2DEG and conduction band) show hexagonal contours near the Fermi level (435~ meV above the Dirac point). This hexagonal warping of the topological state is apparent from its different Fermi wavevectors of $k_F= 0.134(1)$ and 0.128(1)~\AA$^{-1}$ along the $\bar{M}\bar{\Gamma}\bar{M}$ and  $\bar{K}\bar{\Gamma}\bar{K}$ directions, respectively. Similar deviations from an ideal Dirac cone shape have already been described for the case of Bi$_2$Te$_3$  \cite{Chen:2009,Fu:2009} where the effect  is more pronounced than here. For the conduction band and the 2DEG, the hexagonal shape implies a non-isotropic effective mass, a point we shall return to later.

It is important to note the very different physical nature of the topological state and the 2DEG. The former is completely non spin-degenerate, apart from the Dirac point. As such, its quasiparticles are protected from back-scattering. The band structure does not permit the opening of energy gaps due to a small perturbation, or the formation of metal-insulator transitions in the form of charge density waves, despite the strongly nested Fermi surface \cite{Kim:2005b} (a spin-density wave would be permitted \cite{Fu:2009}). The 2DEG, on the other hand, is spin degenerate and could be prone to such instabilities. In fact, a hexagonal Fermi surface with its strong nesting has, to the best of our knowledge, not yet been observed for such a state.

Most importantly, the existence of the 2DEG opens new decay channels for the topological state. An indication of this is the change in the linewidth of the latter. Immediately after cleaving the crystal, the topological state is found to be very narrow, of the order of $\Delta k \approx 0.016(5)$~\AA$^{-1}$ at the Fermi level. If we interpret this width as a measure for the inverse mean free path on the surface $l$, this would correspond to $l \approx 70~$\AA  \cite{Kevan:1986}. This value is similar to what is found for highly perfect epitaxial graphene \cite{Bostwick:2007} or cleaved high-temperature superconductors \cite{Borisenko:2006a}. A long mean free path for the topological state is, of course, expected from the reduced phase space for scattering \cite{Nechaev:2009}. Once the 2DEG is established at the surface, the width of the topological state is increased to  $\Delta k \approx 0.019(5)$~\AA$^{-1}$, while the 2DEG state is slightly wider ($\Delta k \approx 0.024(5)$~\AA$^{-1}$). Apart from the presumably small increased number of defects near the surface, the increased width of the topological state is almost certainly related to the presence of the 2DEG itself, and the scattering processes now permitted between this and the topological state.

\begin{figure}
\begin{center}
\includegraphics[width=0.9\textwidth]{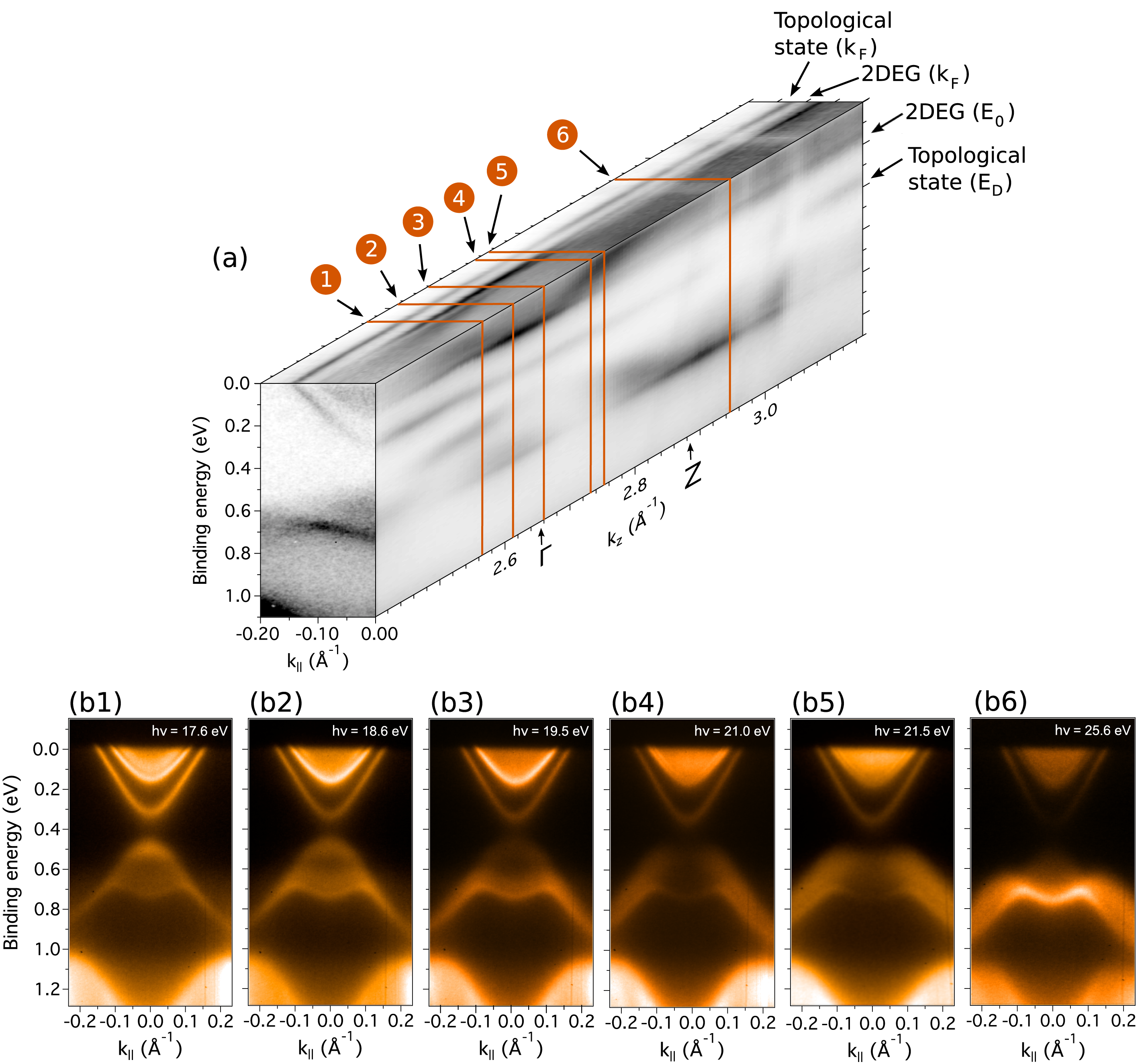}
\caption{{\bf $k_z$-dependence of the electronic structure of Bi$_2$Se$_3$.} (a) Three-dimensional representation of the electronic structure evolution of Bi$_2$Se$_3$ with $k_z$, determined from ARPES measurements taken at 181 photon energies between 14 and 32~eV, between 7 and 11 hours after cleaving the sample.  The bulk $\Gamma$ and Z high symmetry points are indicated. Selected cuts (marked 1--6) are shown, measured at photon energies of (b1) 17.6, (b2) 18.6, (b3) 19.5, (b4) 21.0, (b5) 21.5, and (b6) 25.6 eV. The measurements are taken at an angle enclosing 10.5~$^{\circ}$ with the $\bar{K}\bar{\Gamma}\bar{K}$ direction. The complete data set is available in the online supplementary material of this paper. Apart from intensity modulations due to matrix element effects, the topological state and the 2DEG do not change with photon energy (b1--6), leading to marked linear features in (a) on the $k_{\parallel}-k_z$ (at the Fermi wavevectors, $k_F$) and the $E_b-k_z$ planes (at the Dirac point, E$_D$, and 2DEG band bottom, E$_0$, respectively). This lack of $k_z$ dispersion confirms their two-dimensional nature. A third linear feature in the $E_b-k_z$ plane ($E_b\approx750$~meV) is a surface state in the projected bulk band gap below the upper valence band (M-shaped feature, clearly visible in (b6)).
 \label{fig:2}}
\end{center}
\end{figure}

The dimensionality of all of the observed states can be probed directly by performing ARPES measurements at different photon energies, thus varying the wave vector perpendicular to the surface, $k_z$. Fig. \ref{fig:2}(a) shows the result of such measurements, revealing the evolution of the electronic states on the Fermi surface, and within 1.1~eV of the Fermi level at $\bar{\Gamma}$, with $k_z$. Selected measurements at different photon energies are also shown in Fig. \ref{fig:2}(b) (the full data set is available in the supplementary material). Both the topological state and the 2DEG do not disperse with $k_z$, confirming their two-dimensional nature: they occur at the same energy and momentum in Figs. \ref{fig:2}(b1--6) and give rise to linear features on both the $k_{\parallel}-k_z$ and the $E_b-k_z$ planes (Fig. \ref{fig:2}(a)). In contrast, a strong $k_z$-dispersion of the bulk conduction (valence) band is evident, showing a maximum (minimum) in binding energy at the bulk $\Gamma$ point at $k_z \approx 2.65$~\AA$^{-1}$.  This dispersion also explains the photon-energy-dependent ``filling'' of the 2DEG contour, due to emission from bulk conduction band states (Fig. \ref{fig:2}(b1) and (b3)). 

While the topological and 2DEG states do not disperse in $k_z$, these spectral features do show strong intensity variations with photon energy. Such a $k_z$-dependence of the photoemission intensity of two-dimensional states is well known for metallic surface states, due to transition matrix element effects \cite{Louie:1980,Hofmann:2002}. From Fig. \ref{fig:2}(a), emission from the 2DEG is resonantly enhanced when the conduction band reaches its minimum at the bulk $\Gamma$ point. In the $E_b-k_z$ plane (Fig. \ref{fig:2}(a)), another state is observed at a binding energy of $\approx$750~meV which does not disperse with $k_z$. This corresponds to the M-shaped feature in the individual cuts, particularly pronounced in Fig. \ref{fig:2}(b6). We interpret this as another surface state, situated in a projected bulk band gap below the upper valence band near $\bar{\Gamma}$ \cite{Xia:2009}. Similar to the 2DEG, emission from this state is resonantly enhanced when the bulk valence band dispersion reaches a minimum at the Z point ($k_z \approx$~2.9~\AA$^{-1}$).

\begin{figure}
\begin{center}
\includegraphics[width=0.8\textwidth]{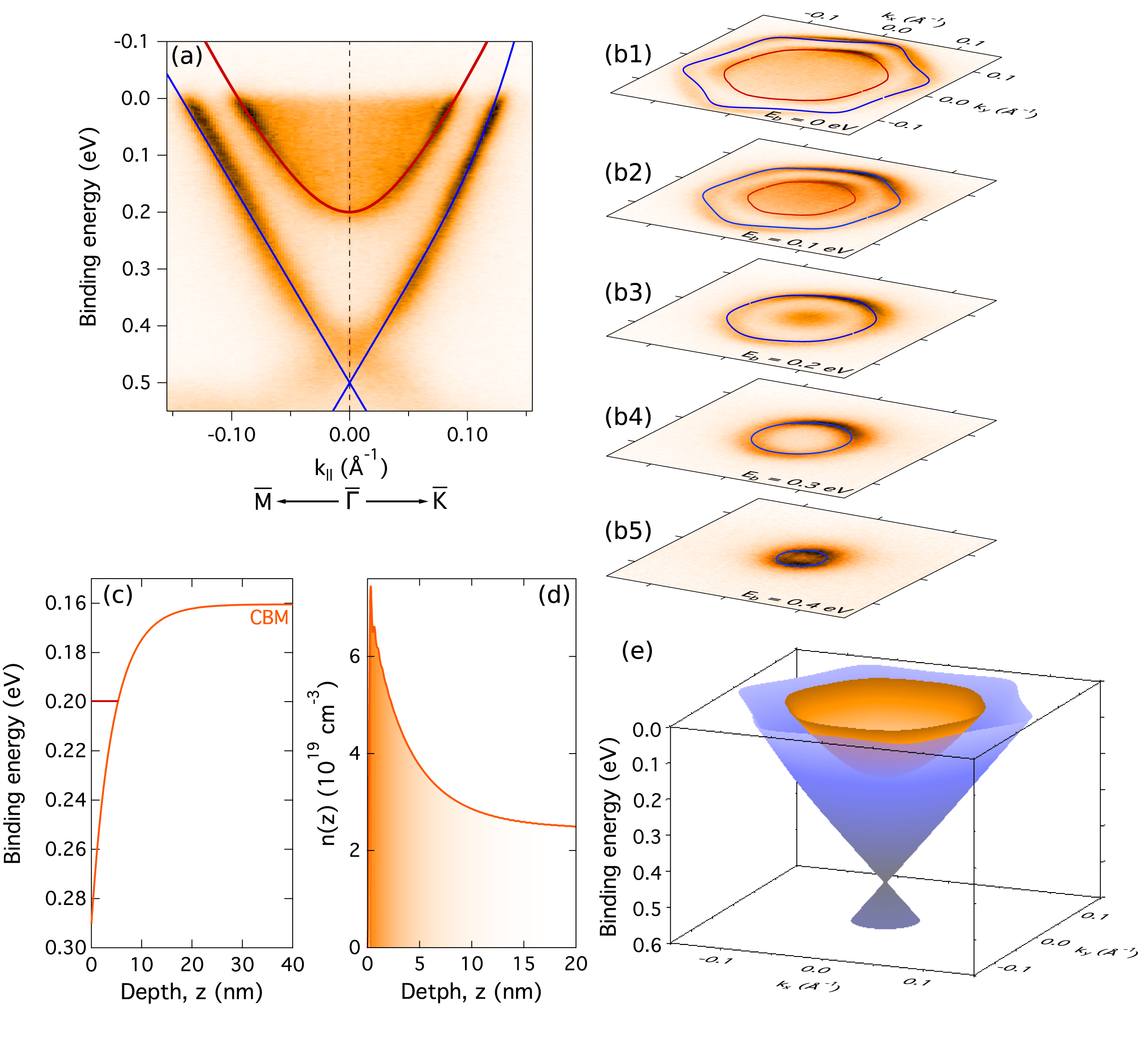}
\caption{{\bf Model calculations for the 2DEG and topological state.} Coupled Poisson-Schr{\"o}dinger calculations, incorporating conduction-band non-parabolicity and anisotropy of the effective mass, well reproduce the (a) dispersion and (b) constant energy contours of the 2DEG state. The downward bending of the conduction band at the surface and corresponding increase in near-surface carrier density (that is, the 2DEG formation) are shown in (c) and (d), respectively. The topological state is also reasonably characterized by model calculations describing hexagonal warping of the linear Dirac dispersion (a,b). The co-existence of the 2DEG and topological surface state is represented in (e).
 \label{fig:3}}
\end{center}
\end{figure}

The 2DEG on the surface is derived from the conduction band states, as indicated by its resonant enhancement near $\Gamma$, but confined in the $z$-direction due to a strong band bending near the surface. 
We have performed model calculations to further substantiate the existence and origins of this 2DEG, as shown in Fig. \ref{fig:3}. The near-surface band bending of the material and the dispersions of the resulting 2DEG states are calculated via a coupled solution of the Poisson and Schr{\"o}dinger equations, based around the method described in Ref. \cite{King:2008}. These calculations accurately reproduce the measured dispersion and constant energy contours of the 2DEG state, shown in Fig. \ref{fig:3}(a,b), for a downward band bending at the surface of 0.13 eV (Fig. \ref{fig:3}(c)), very similar to the shift of the Dirac point between the freshly cleaved sample and following development of the 2DEG. It is this band bending which creates a potential well at the surface, causing quantum-confinement of the conduction band states, and development of the 2DEG. In the presence of a potential gradient, one may expect to observe a Rashba spin-splitting of the 2DEG state. However, from a simple model using the calculated band bending potential gradient, we estimate an upper limit for the Rashba splitting of only $\sim\!0.004$~\AA${^{-1}}$ at the Fermi level,  unlikely to be observable in practice.

In addition to the main 2DEG state located $\sim\!40$ meV below the conduction band edge of the bulk (shown in Fig. \ref{fig:3}(c), and clearly apparent in the measured data (Fig. \ref{fig:3}(a)), a second 2DEG state is predicted in the calculations at only $\sim\!7$ meV below the bulk band edge. This state is not shown in Fig. \ref{fig:3} because, being so shallow, its wavefunctions  will be very extended in $k_z$: consequently, we would expect this state to be very delocalized and to largely contribute to the diffuse intensity above the lower 2DEG state. However, we note that at some photon energies, a second state is discernible above the 2DEG band bottom (for example, Figs. \ref{fig:2}(b4,5)). This second state is situated approximately 35~meV above the band bottom of the main 2DEG state, and so would be consistent with a second very shallow state as predicted by the calculations.

The hexagonal Fermi surface contour of the 2DEG state implies an anisotropic in-plane effective mass, and we find values of 0.122m$_0$ and 0.109m$_0$ along $\bar{\Gamma}\bar{M}$ and $\bar{\Gamma}\bar{X}$, respectively. The 2DEG state also exhibits some deviations from parabolicity. We attribute this to a $\mathbf{k}\cdot\mathbf{p}$ interaction~\cite{Kane:1957} between the conduction and valence bands, which we include within our calculations. A $\mathbf{k}\cdot\mathbf{p}$ model has also been proposed to describe the hexagonal warping of the topological surface state in Bi$_2$Te$_3$  \cite{Fu:2009}. This model reproduces the dispersions and constant energy contours of the topological state observed here reasonably well (Fig. \ref{fig:3}(a,b)), although the magnitude of the hexagonal warping term is approximately a factor of 2.3 smaller than that estimated from the measured Fermi velocity and wavevector, using the formalism within Ref. \cite{Fu:2009}.

The model calculations confirm the experimental findings presented above that a downward band bending at the surface of Bi$_2$Se$_3$ induces a two-dimensional electron gas, that co-exists with the topological surface state (Fig. \ref{fig:3}(e)). This suggests the potential to modulate the 2DEG density via control of the band bending, either through absorption of elemental or molecular species on the clean surface, or via a gate electrode. Such a scheme could potentially be used to obtain a switchable metallic, and possibly superconducting, layer in the vicinity of the topological surface state, which could be fundamental to the incorporation of the novel properties of topological insulators into device applications.

\textbf{Methods Summary} 
Single crystals of Bi$_2$Se$_3$ were grown as follows: Stoichiometric mixtures of 5N purity elemental Bi and Se were melted at 860 $^{\circ}$C for 24 hours in an evacuated quartz ampoule, cooled down to 650  $^{\circ}$C at a rate of 2.5  $^{\circ}$C/h, and then annealed at 650  $^{\circ}$C for another two days. The sample was then removed to another evacuated quartz ampoule with a conical bottom and zone melted through an induction coil with a rate of 1.2 mm/h. 

Angle-resolved photoemission measurements were performed at the SGM-3 beamline of the ASTRID synchrotron radiation facility in Aarhus. The samples were cleaved at room temperature and ARPES data were taken at around 60~K.  Between cleaving and the reported measurements, the sample temperature was thus above 60~K,  such that the more rapid change in the Fermi energy compared to that reported by Hsieh \emph{et al.} \cite{Hsieh:2009c} can be explained by the higher sample temperature. For the detailed mapping of the states, the energy and angular resolutions were set to better than 10~meV and 0.13$^{\circ}$, respectively.  The photon energy for the data in Figs. \ref{fig:1} and \ref{fig:3} was 16~eV. For the photon energy scan in Fig. \ref{fig:2}, the energy resolution was better than 20~meV. 
The $k_z$ values for the three-dimensional representation of the photon energy scan of Fig. \ref{fig:2} were calculated using free-electron final states, i.e. $k_z=\sqrt{2m_e / \hbar}(V_0 + E_{\mathrm{kin}} \cos^2 \Theta)^{1/2}$, where $\Theta$ is the emission angle and $V_0$ is the inner potential, chosen to be 11.8~eV, in good agreement with Ref. \cite{Xia:2009}. The intensity of each individual image has been normalized such that the maximum amount of detail is  visible in the $k_z$ maps. 

The coupled Poisson-Schr{\"o}dinger calculations \cite{King:2008} were performed incorporating both the non-parabolicity of the conduction band dispersion as well as the anisotropy of the effective mass along different crystallographic directions. The in-plane mass was taken as 0.122m$_0$ and 0.109m$_0$ along the $\bar{\Gamma}\bar{M}$ and $\bar{\Gamma}\bar{X}$ directions, respectively, where m$_0$ is the free-electron mass. Along k$_z$, an effective mass of 0.24m$_0$ was used, estimated from the bulk conduction band dispersion shown in Fig. \ref{fig:2}(a). The band gap was taken as 350 meV, and a bulk dielectric constant of 113 was used. The bulk carrier density was set at $2.45\times10^{19}$ cm$^{-3}$, corresponding to a bulk Fermi level 0.16 eV above the conduction band edge, as estimated from the ARPES data.

\textbf{Acknowledgements}
Financial support from the Danish Council for Independent Research and the Danish National Research Foundation is gratefully acknowledged. 



\end{document}